\documentclass[twocolumn]{aastex61}
 
\usepackage{natbib}
\usepackage{enumerate}
\usepackage{hyperref}
\usepackage{color,soul}

\shorttitle{The Monoceros Ring and A13}
\shortauthors{Sheffield et al.}

\begin{document}
\title{A Disk Origin for the Monoceros Ring and A13 Stellar Overdensities}

\author{Allyson~A.~Sheffield}
\affiliation{Department of Natural Science,
City University of New York, LaGuardia Community College, Long Island City, NY 11101, USA}

\author{Adrian~M.~Price-Whelan}
\affiliation{Department of Astrophysical Sciences,
Princeton University, Princeton, NJ 08544, USA}

\author{Anastasios Tzanidakis}
\affiliation{Department of Astronomy, Columbia University, Mail Code 5246,
New York, NY 10027, USA}

\author{Kathryn~V.~Johnston}
\affiliation{Department of Astronomy, Columbia University, Mail Code 5246,
New York, NY 10027, USA}

\author{Chervin~F.~P.~Laporte}
\affiliation{Department of Astronomy, Columbia University, Mail Code 5246,
New York, NY 10027, USA}

\author{Branimir Sesar}
\affiliation{Max Planck Institute for Astronomy, K{\"o}enigstuhl 17, 69117, Heidelberg, Germany}

\correspondingauthor{Allyson~A.~Sheffield}
\email{asheffield@lagcc.cuny.edu}

\begin{abstract}
The Monoceros Ring (also known as the Galactic Anticenter Stellar Structure) and A13 are stellar overdensities at estimated heliocentric distances of $d \sim 11$ kpc and 15 kpc observed at low Galactic latitudes towards the anticenter of our Galaxy. While these overdensities were initially thought to be remnants of a tidally-disrupted satellite galaxy, an alternate scenario is that they are composed of stars from the Milky Way (MW) disk kicked out to their current location due to interactions between a satellite galaxy and the disk. 
To test this scenario, we study the stellar populations of the Monoceros Ring and A13 by measuring the number of RR Lyrae and M giant stars associated with these overdensities. We obtain low-resolution spectroscopy for RR Lyrae stars in the two structures and measure radial velocities to compare with previously measured velocities for M giant stars in the regions of the Monoceros Ring and A13, to assess the fraction of RR Lyrae to M giant stars ($f_{RR:MG}$) in A13 and Mon/GASS. 
We perform velocity modeling on 153 RR Lyrae stars (116 in the Monoceros Ring and 37 in A13) and find that both structures have very low $f_{RR:MG}$. The results support a scenario in which stars in A13 and Mon/GASS formed in the MW disk. We discuss a possible association between Mon/GASS, A13, and the Triangulum-Andromeda overdensity based on their similar velocity distributions and $f_{RR:MG}$.
\end{abstract}


\section{Introduction}
The Milky Way is complex and dynamic. The density profile of the halo is not smooth (e.g., \citealt{juric08}) but rather populated by substructures, in the form of stellar streams and less collimated overdensities. Photometric detections of stellar streams and other substructures \citep{mswo, belokurov06, bell08} are consistent with predictions from cosmological simulations that link stellar streams to accreted dwarf satellite galaxies (e.g., \citealt{bj05}, \citealt{bell08}, \citealt{johnston08}).  
While this comparison suggests that the majority of stars in the Galactic halo were accreted, some fraction of halo stars should have formed in situ, either at their current location (as part of the dissipative collapse that formed the Galaxy; \citealt{els62}) or deep in the Milky Way's potential well and subsequently relocated to the halo (kicked-out) due to a dynamical perturbation to the disk. This leads to the key question of whether there are any in situ stars in our halo.

Chemical tagging of stars in dwarf galaxies (\citealt{fb02}, \citealt{venn04}) provides a window into the environment in which the stars formed, as stars maintain the chemical imprint of their primordial gas cloud. In this way, stars that originated in the Milky Way (MW) can be separated from accreted satellite stars. Recently, \citet{bonaca17} studied the chemical and dynamical properties of local halo stars with heliocentric distances $<$ 3 kpc selected from the Tycho-Gaia astrometric solution (TGAS; \citealt{tgas}) combined with the Apache Point Observatory Galactic Evolution Experiment (APOGEE; \citealt{apogee}) and the Radial Velocity Experiment (RAVE; \citealt{steinmetz06}) spectroscopic surveys. Stars were selected kinematically to have speeds $>$ 220 km s$^{-1}$ relative to the Local Standard of Rest, and Galactic components were assigned using a Toomre diagram; chemically, many of the stars in the halo region of the Toomre diagram are metal-rich. The metallicity distribution for the halo stars is bimodal, with one peak at [Fe/H] $\sim$ -0.5 and a tail extending out to a metallicity of roughly -1; the stars in this metal-rich distribution also fall along the trend for MW disk stars in the [$\alpha$/Fe]-[Fe/H] plane. The results of \citet{bonaca17} provide compelling evidence for a local population of in situ halo stars with disk-like abundances but halo-like kinematics.

There has been recent interest in other evidence for kicked-out material found around the outskirts of the MW (see \citealt{johnston17} for a review of results from recent observations and simulations). For example, \citet{pw15} (hereafter PW15) used stellar populations -- specifically, the fraction of RR Lyrae stars to M giants $f_{RR:MG}$ -- to determine the origin of the diffuse Triangulum-Andromeda (TriAnd) cloud \citep{majewski04,rp04,martin07,sheffield14,perottoni17}. Using a mixture model to represent the velocity distribution of stars in the structure and halo, PW15 found $f_{RR:MG}$ for TriAnd consistent with a kicked-out population. Using a mixture model to represent the velocity distribution of stars in the structure and halo, PW15 estimated $f_{RR:MG}$ for TriAnd to be $<$ 0.38 (at 95\% confidence). In contrast, using studies in the literature, they estimated $f_{RR:MG}$ $\sim$ 0.5 for the LMC and Sgr, and $>>$ 1 for smaller satellites of the MW which all have RRL populations, but no M giants. They concluded that TriAnd was consistent with a kicked-out population.

The Monoceros Ring (also known as the Galactic Anticenter Stellar Structure, GASS, hereafter referred to as Mon/GASS) is another example of a low-latitude ``ring-like'' stellar feature wrapping around the Galactic midplane \citep{slater14,morganson16}, first detected with Sloan Digital Sky Survey (SDSS; \citealt{sdss}) A-F dwarfs and main sequence turnoff (MSTO) stars  \citep{yanny00, newberg02}. Moreover, Mon/GASS is spatially continuous with A13, which is a stellar overdensity in M giants detected using the EnLink group-finding algorithm applied to stars in 2MASS \citep{sharma10}. 
A detailed spectroscopic analysis of A13 was carried out by \citet{li17} who found that the radial velocities of A13 M giants have a cold dispersion, inconsistent with that expected for a pressure-supported distribution of halo giants. \citet{li17} also looked at connections between A13, Mon/GASS, and TriAnd. They compared the radial velocities in the Galactic Standard of Rest (GSR) frame for all three structures and found a continuous sequence as a function of Galactic longitude. Considering the spatial differences and kinematic similarities of these features, one possibility is that they all originated in the disk, but were heated to their current locations in the halo via a dynamical perturbation to the disk. Further evidence for a disk origin of the Monoceros Ring comes from the recent work of \citet{deason18}. Using proper motions from the SDSS-Gaia source catalog and metallicities from Segue, the Monoceros Ring has kinematical and chemical properties consistent with the thin disk; the Eastern Banded Structure (EBS), a stellar feature first detected by \citep{grillmair06,grillmair11} at $\sim$ 10 kpc with Hydra I as a possible progenitor, shows evidence of a connection to the Monoceros Ring at the disk-halo interface. \citet{deboer18} also used SDSS-Gaia proper motions and concluded that the Monoceros Ring may be due to a perturbation to the disk, based on tangential motions.

\begin{figure*}[ht]
\begin{centering}
\includegraphics[scale=0.5]{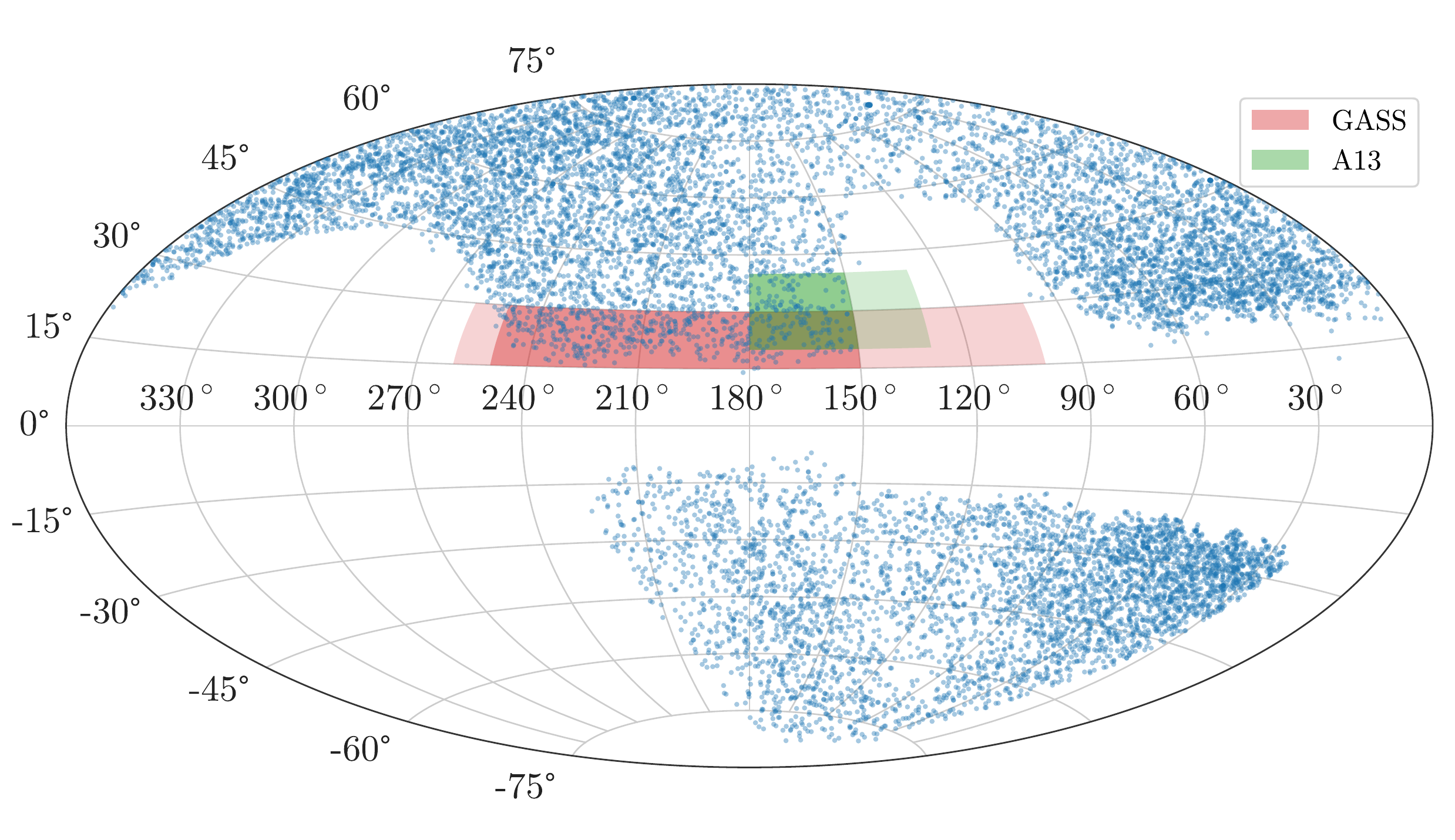}
\caption{Selection sub-regions for the program RR Lyrae stars from the Catalina Sky Survey (CSS) footprint. The lightly shaded areas are the M giant regions, and the dark shaded areas are the effective sky regions of the structures due to the CSS footprint.}
\label{css}
\end{centering}
\end{figure*}

As with the halo, the MW's disk is not a quiescent component.  
\citet{gilmore83}, analyzing stars in the direction of the South Galactic Pole, found that the density profile of the disk is best fit by two exponentials, with scale heights of 300 pc and 900 pc for the thin and thick disks respectively. 
Recent work has shed new light on the radial and vertical scales of the disk components. \citet{widrow12} used main sequence stars in SDSS DR8 to calculate number counts of stars in the Northern and Southern Galactic hemispheres close to the plane; they found that there is a North-South asymmetry in the stellar number counts and this was verified in simulations \citep{gomez13}. A similar asymmetry in the bulk vertical and radial motions was found by \citet{carlin13} using Large Sky Area Multi-Object Fiber Spectroscopic Telescope (LAMOST; \citealt{cui12}) spectra combined with PPMXL \citep{roeser10} proper motions. This kinematic asymmetry was further verified by \citet{williams13}, who used spectra from RAVE. 
These results support models which show that a fairly massive dwarf satellite \citep{younger08} or multiple mergers \citep{kazan08} can induce such asymmetries.
\citet{xu15} analyzed F-type MSTO in SDSS and, by looking at differences in the color-magnitude diagrams in the direction of the Galactic anticenter in the North and South, found that there are overdensities that oscillate vertically in $Z$. \citet{antoja17}, using TGAS, detected an oscillating asymmetry in transverse motions as a function of Galactic longitude over the entire sky, close to the midplane; in the direction of the outer disk ($|l|$ $<$ 70$^{\circ}$), the asymmetry is present for all stellar types, providing evidence for a common origin for these features.  

In this work, we aim to clarify the origins of Mon/GASS and A13 by looking at their stellar populations. We took low-resolution ($R \sim$ 2000) spectra of RR Lyrae stars in the same regions of the Galaxy as the M giants in these two overdensities. In section $\S{\ref{data}}$, we discuss the data selection, observations, and reductions. The fraction of RR Lyrae stars to M giants is discussed in $\S{\ref{pops}}$. In $\S{\ref{interp}}$, we present an interpretation of the results, in particular the potential connection between Mon/GASS and A13 with vertical oscillations detected in the Galactic disk. We summarize our findings in $\S{\ref{summary}}$.

\section{\label{data}Data: Observations and Reductions}
\subsection{RR Lyrae selection and observations}
As in PW15, the radial velocities (RVs) of M giant and RR Lyrae stars were collectively analyzed. The M giant RVs for Mon/GASS are taken from \citet{crane03} and the A13 M giants from \citet{li17}. The RR Lyrae stars (only RRab-type) were selected from the Catalina Sky Survey (CSS; \citealt{larson03}) in regions that are coincident with the Mon/GASS and A13 M giant overdensities: for Mon/GASS, $150^{\circ} < l < 260^{\circ}$, $17^{\circ} < b < 30^{\circ}$ and heliocentric distances $7 < d < 15$ kpc; for A13,  $150^{\circ} < l < 180^{\circ}$, $20^{\circ} < b < 40^{\circ}$ and distances $11 < d < 33$ kpc. 
The CSS footprint does not evenly cover, and is some cases provides no coverage of, the regions that were analyzed for the M giants. In Figure \ref{css}, the M giant selection regions are shown along with the RR Lyrae sub-regions.

\begin{figure}[ht]
\begin{centering}
\includegraphics[scale=0.65]{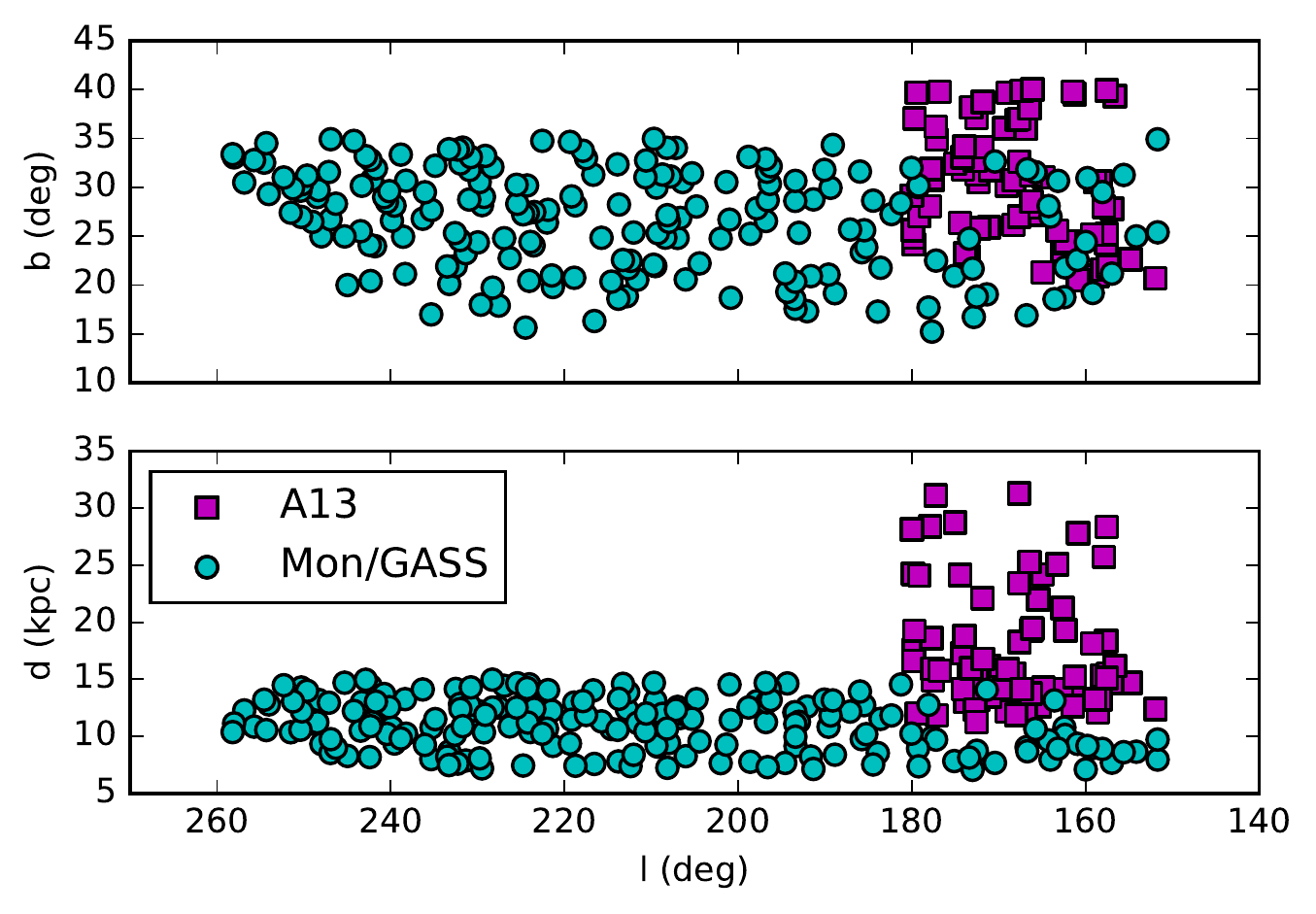}
\caption{Spatial distribution of the Mon/GASS and A13 RR Lyrae stars selected from the Catalina Sky Survey. The top panel shows the Galactic longitude and latitude of the RR Lyrae stars, and the lower panel shows the heliocentric distance as a function of Galactic longitude.}
\label{lbd}
\end{centering}
\end{figure}

This selection yielded 199 Mon/GASS and 72 A13 target RR Lyrae stars. 
The distributions in spatial coordinates and distances are shown in Figure \ref{lbd}.
Of these targets, a total of 153 were observed from 05 February 2016 to 11 February 2016 using the 2.4-m Hiltner telescope at MDM Observatory with ModSpec, equipped with a 600 l mm$^{-1}$ grating, and MDM's SITe \textit{echelle} CCD. 
Each target had one exposure taken, which ranged from 600 to 900 seconds, depending upon the seeing and the brightness (at the phase being observed) of the RR Lyrae star. The seeing ranged from about 1.25'' on the clearest night, to 2.5''. The S/N at the center of the continuum (around 5220 \AA) varies from 20 - 40 for the RR Lyrae targets.

\begin{deluxetable*}{l c c c c c c c c}[t]
\tablewidth{0pt}
\tabletypesize{\scriptsize}
\tablehead{Name$^{a}$ & $l$ & $b$ & Distance$^{b}$ & $v_{hel}$ & $v_{GSR}$ & $v_{err}$ & Observed Phase & Structure \\
      & deg & deg & kpc & km s$^{-1}$  & km s$^{-1}$ & km s$^{-1}$ &  & \\
}      
\tablecaption{\label{photdata}Properties of the observed RR Lyrae stars.}
\startdata
J064339.2+425035	&	172.856076	&	16.732728	&	7.5	&	-141.7	&	-111.1	&	13.9	&	0.72	&	Mon/GASS	\\
J080730.6+302632	&	191.325151	&	28.755251	&	7.1	&	-32.9	&	-73.1	&	14.5	&	0.52	&	Mon/GASS	\\
J080902.8+095641	&	212.654276	&	21.713061	&	13.9	&	210.1	&	87.5	&	13.2	&	0.11	&	Mon/GASS	\\
J080939.2+484725	&	170.425727	&	32.634859	&	7.7	&	-54.2	&	-16.4	&	16.6	&	0.35	&	Mon/GASS	\\
J081359.9+454157	&	174.175330	&	33.055364	&	17.3	&	15.3	&	39.5	&	17.5	&	0.09	&	A13	\\
\enddata
\tablecomments{This table is available in its entirety in a machine-readable form online. A portion is shown here for guidance regarding its form and content.}
\begin{enumerate}[a]
\item Position ID from the Catalina Sky Survey (CSS).
\item Heliocentric distance from CSS.
\end{enumerate}
\end{deluxetable*}

\subsection{Radial Velocities}
The RR Lyrae spectra were reduced using standard \textit{IRAF} processing tasks via PyRAF\footnote{\url{http://www.stsci.edu/institute/software_hardware/pyraf}}. One-dimensional spectra were converted from pixel to wavelength space using \textit{dispcor}. To check the accuracy of the dispersion solution, we looked at the positions of [OI] night sky emission lines (e.g., 5577.3 \AA, 6300.0 \AA); the scatter in this value defines the zero-point velocity error and is $\sim$ 5 km s$^{-1}$. The \textit{rvcorr} task was used to correct for barycentric motion. Heliocentric RVs were found individually for the H$_{\alpha}$ and H$_{\beta}$ lines. This is necessary as the H Balmer lines form at different layers in the stellar atmosphere of an RR Lyrae star, thus leading to differences in the RVs as a function of phase.

To determine the heliocentric RV, the line profiles for each Balmer line, H$_{\alpha}$ and H$_{\beta}$, were separately convolved with an antisymmetric function (the derivative of the Gaussian function), as described in \citet{sy80}. The line center is taken as the zero of the convolution. The uncertainties are calculated from the variance of the extracted 1D spectra. The RVs were also calculated using the pixel-fitting methodology described in \citet{koposov11}, using the spectral templates from \citet{munari05} and are in precise agreement. For example, the RR Lyrae target J073645.0+533316 has a derived observed RV of -365 km s$^{-1}$ from the former methodology and -368 km s$^{-1}$ from the latter. 

The methodology described in \citet{sesar12} was used to compute the systemic (center-of-mass) velocities for the RR Lyrae stars. The spectra cover a range from $\sim$ 4000 -- 7200 \AA. Although this includes the H$_{\delta}$ line at 4102 \AA\ and the H$_{\gamma}$ line at 4341 \AA, these lines are in the low S/N region of the spectra and therefore were not used. The RR Lyrae stars were typically observed at one phase; for a check on the methodology, we measured spectra at two phase points for 3 stars. For a given phase point, the heliocentric H$_{\alpha}$ and H$_{\beta}$ RVs were individually fit to the H$_{\alpha}$ and H$_{\beta}$ RV curves from \citet{sesar12}. The variances of the H$_{\alpha}$ and H$_{\beta}$ derived systemic velocities, $\sigma_{\alpha}^2$ and $\sigma_{\beta}^2$, include the uncertainties in the heliocentric RVs for each of the Balmer lines. As in \citet{sesar13}, the systemic velocities and total error are weighted by the inverse of the variances. The systemic velocities were converted from the heliocentric frame to the GSR frame, and we adopted the values $\Theta_{0}$ = 236 km s$^{-1}$ \citep{bovy09} and ($U_{\sun},V_{\sun},W_{\sun}$) = (11.1, 12.2, 7.3) km s$^{-1}$ \citep{schoenrich10} to correct for solar motion. The derived systemic velocities of program RR Lyrae stars are presented in Table \ref{photdata}.

The systemic velocities are found at phase point $\phi$ = 0.27. The phase of the systemic velocity is the phase at which the pulsation velocity is equal to 0; this happens when the atmosphere is most extended and at its turn-around point. During one pulsation cycle, the atmosphere travels the distance equal to 2$\Delta R$, where $\Delta R$ is the maximum extent of the atmosphere (i.e., the atmosphere extends by $\Delta$R and then it contracts by $\Delta R$). 2$\Delta R$ can be calculated using the radial velocity curve:

\begin{equation}
2\Delta R = \int_{0}^{1} \rm RV(\phi) d\phi
\end{equation}

The systemic velocity, $\phi_{\rm sys}$, is then the phase for which

\begin{equation}
\Delta R = \int_{0}^{\phi_{\rm sys}} \rm RV(\phi) d\phi
\end{equation}

or

\begin{equation}
\frac{1}{2} =  \frac{\int_{0}^{\phi_{\rm sys}} \rm RV(\phi) d\phi}{\int_{0}^{1} \rm RV(\phi) d\phi}
\end{equation}

If the above approach is applied to the Balmer RV curves, one gets $\phi_{\rm sys} \sim$ 0.27, which is very close to the systemic phase measured from metallic lines in the study by \citet{kolenberg10}.

\subsection{Density of RR Lyrae Stars}
\begin{figure}[ht]
\begin{centering}
\includegraphics[scale=0.35]{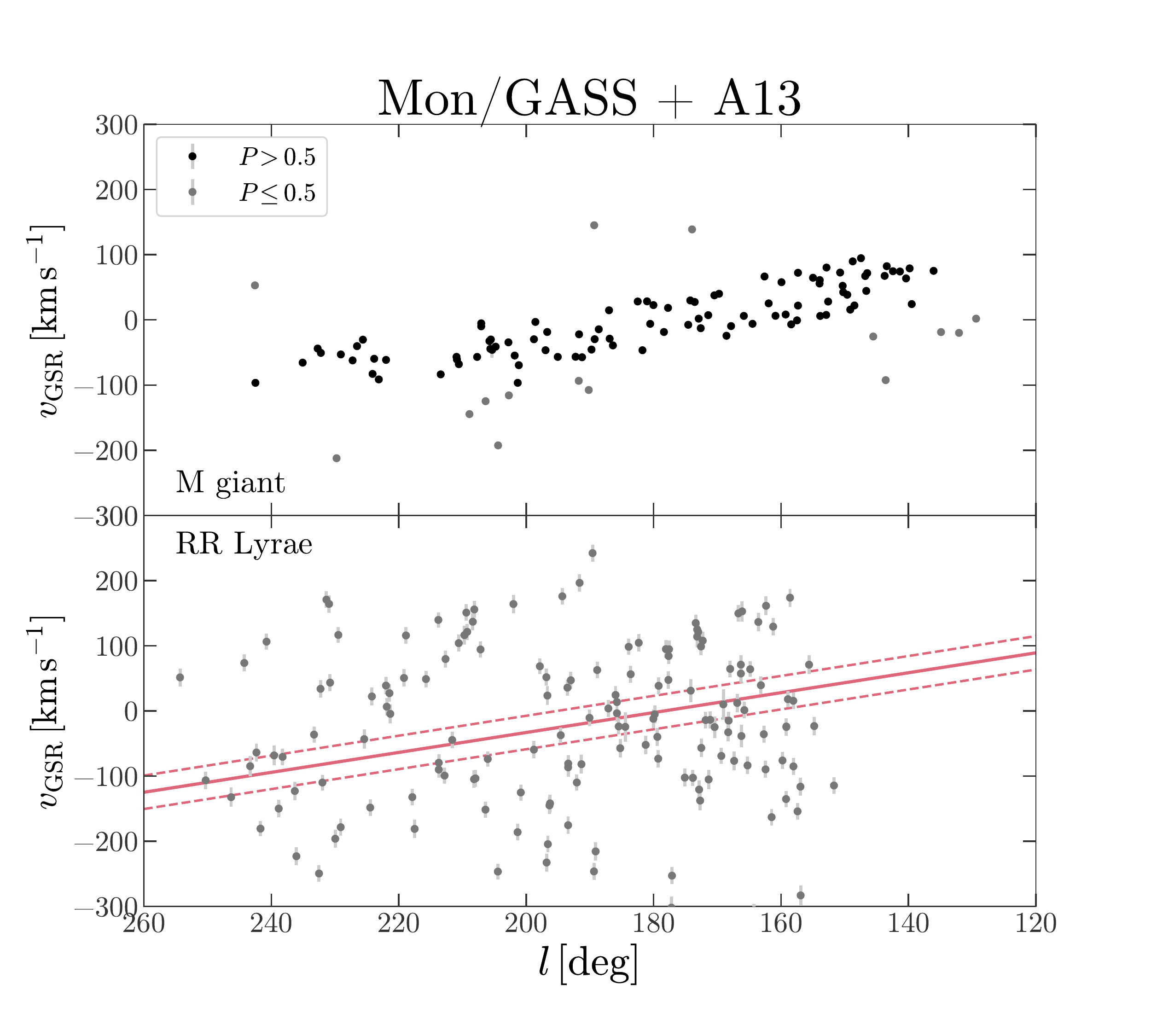}
\caption{Distribution of $v_{\rm GSR}$ as a function of Galactic longitude for Mon/GASS+A13 M giants (top panel) and RR Lyrae stars (bottom panel). Black points denote stars with a posterior probability greater than 0.5 of membership in a cold velocity sequence. In the lower panel, the MAP fit to the M giant velocity sequence is shown as the solid red line, and the dashed red lines show the MAP derived dispersions in the velocity sequence.}
\label{lvgsr_mon_gass_a13}
\end{centering}
\end{figure}

To determine the completeness of the Mon/GASS and A13 RR Lyrae targets selected from CSS, a comparison was made with RR Lyrae stars from Pan-STARRS1 (PS1). Within 40 kpc, the PS1 RR Lyrae star selection is 92\% complete \citep{sesar17a}. The PS1 and CSS RR Lyrae stars were binned by distance and the ratio of the PS1 to CSS counts was found for each bin, accounting for the 92\% completeness of the PS1 selection. This yielded a PS1 completeness corrected number of RR Lyrae in GASS, within the range 7-15 kpc, of 299; for A13, in the range 11-33 kpc, the completeness corrected number is 101.

\begin{figure}[ht]
\begin{centering}
\includegraphics[scale=0.5]{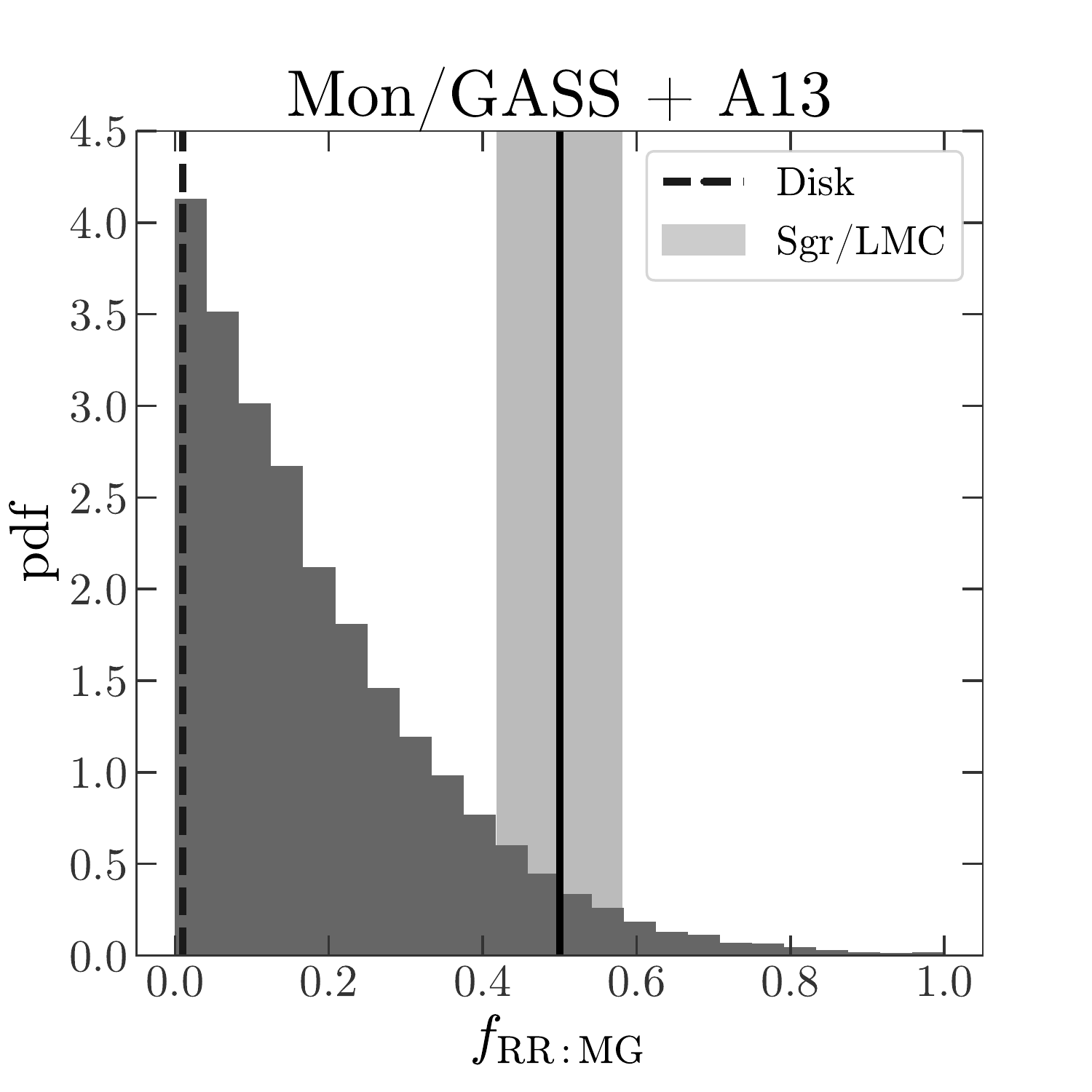}
\caption{Binned samples from the marginal posterior probability over the number ratio of RR Lyrae to M giants in Mon/GASS+A13. The dashed line represents the Galactic disk and the solid line shows fairly massive satellite galaxies (Sgr and the LMC).}
\label{frrmg_mon_gass_a13}
\end{centering}
\end{figure}

Using the smooth halo density relation from \citet{sesar11}, the \textit{expected} number of RR Lyrae stars in the spatial and distance ranges listed above were found to be 250 for Mon/GASS and 320 for A13. The number of RR Lyrae expected in the vicinity of A13 is overestimated more than three-fold; this is not surprising, as the range selected for A13 stars extends to 33 kpc, a region in the stellar halo that is know to be highly substructured \citep{sesar10}. Although Mon/GASS covers a similar swath of sky as A13, a smaller volume is encompassed so the slight overdensity of RR Lyrae stars is expected -- generally, fewer RR Lyrae stars are found at these closer distances due to their low metallicities.

\section{\label{pops}Stellar Populations: Ratio of RR Lyrae to M giant Stars}
To estimate what fraction of the observed stars are members of the Mon/GASS and A13 stellar features, the same methodology described in Sections 3.2 and 3.3 of PW15 was carried out. In brief, the posterior probability of an individual star belonging to one of the sequences was determined by using a Gaussian mixture model for the M giants and RR Lyrae stars, with one Gaussian for the low dispersion features and the second Gaussian for a high dispersion (random) halo population.
The velocity dispersions of the Mon/GASS and A13 RR Lyrae and M giant stars were compared; 
finding similar dispersions for both types of stars would indicate the presence of a significant population of RR Lyrae in these structures, and lend support for a scenario in which the structures formed in an accretion event.
The comparison was done two ways, (1) for the combined Mon/GASS+A13 sample and (2) for Mon/GASS and A13 separately. 
The results for the combined Mon/GASS+A13 analyses are shown as follows: Figure \ref{lvgsr_mon_gass_a13} shows $v_{\rm GSR}$ as a function of Galactic longitude for the Mon/GASS+A13 M giants and RR Lyrae stars, with the best-fit M giant sequence over-plotted on the RR Lyrae panel and stars with a statistically significant ($P > 0.5$) posterior probability of belonging to the stellar feature marked in black; Figure \ref{frrmg_mon_gass_a13} shows the probability density function of $f_{RR:MG}$, with the results from PW15 for the massive satellites Sagittarius and the LMC included for comparison.

As seen in Figure \ref{lvgsr_mon_gass_a13}, the systemic velocities of the RR Lyrae stars for Mon/GASS+A13 show a large dispersion and all of the RR Lyrae stars have $P < 0.5$. If Mon/GASS and A13 are associated with an accreted dwarf galaxy, then their stellar populations should contain both metal-poor RR Lyrae stars and metal-rich M giants. Figure \ref{frrmg_mon_gass_a13} also shows this: the maximum a posterior (MAP) value of $f_{RR:MG}$ for Mon/GASS+A13 is 0.02 and the MAP value of the velocity dispersion of the combined structure is 25.7 km s$^{-1}$. 
For the separate analyses, the MAP values of $f_{RR:MG}$ are 0.001 for Mon/GASS and 0.03 for A13. The slightly less constraining MAP value of $f_{RR:MG}$ for A13 is likely due to the limited coverage in Galactic longitude. Ultimately, following estimates presented in PW15, all of the $f_{RR:MG}$ MAP values are much more aligned with the disk as opposed to tidally disrupted satellite debris: larger satellites have $f_{RR:MG}$ $\sim$ 0.5 and smaller satellites have $f_{RR:MG}$ $>>$ 1. We have ruled out an accreted origin in favor of a kicked-out origin for both Mon/GASS and A13 (and TriAnd; PW15) by analyzing the stellar populations in these low-latitude overdensities, as evidenced by the lack of \textit{any} RR Lyrae stars with a posterior probability $P > 0.5$ and the low $f_{RR:MG}$ in these features.

\section{\label{interp}Discussion}
\subsection{\label{disksub}Implication for the Origin of Substructure Around the Disk}
In this paper, the origin of the Mon/GASS and A13 stellar features -- accreted or in situ -- was tested by analyzing the stellar populations within these substructures. A satellite piercing the disk perpendicularly can create arc-like features reminiscent of those observed, but tidal debris from an accreted satellite can also produce similar features (e.g., the Sgr tidal streams have multiple wraps, extending over a wide range of distances; \citealt{sesar17b}). Hence, a photometric and kinematic analysis alone cannot provide an unambiguous distinction.

The data collected here demonstrate that the velocities for the RR Lyrae stars observed in both Mon/GASS and A13 have a high dispersion, compared to the M giant velocity sequence, leading to a fraction of RR Lyrae stars to M giant stars in both features that is consistent with a disk population, rather than debris from an accreted satellite galaxy -- as was also found for the TriAnd stellar cloud (PW15). Our results are further supported by recent results from \citet{bergemann18}, which measured the chemical abundances of 6 stars in A13 and 8 stars in TriAnd from high-resolution spectra. An extremely low scatter was found in both [Fe/H] and the other elements analyzed (Na, O, Ti, Eu, and Mg), on the order of 0.06 dex, in both stellar features. All of the abundance ratios (e.g., [Fe/H] = -0.59) are consistent with the disk component of the Galaxy, lending strong support to the kicked-out disk scenario.

\subsection{\label{ring}Connections to the Oscillating Disk?}
\begin{figure}
\begin{centering}
\includegraphics[scale=0.52]{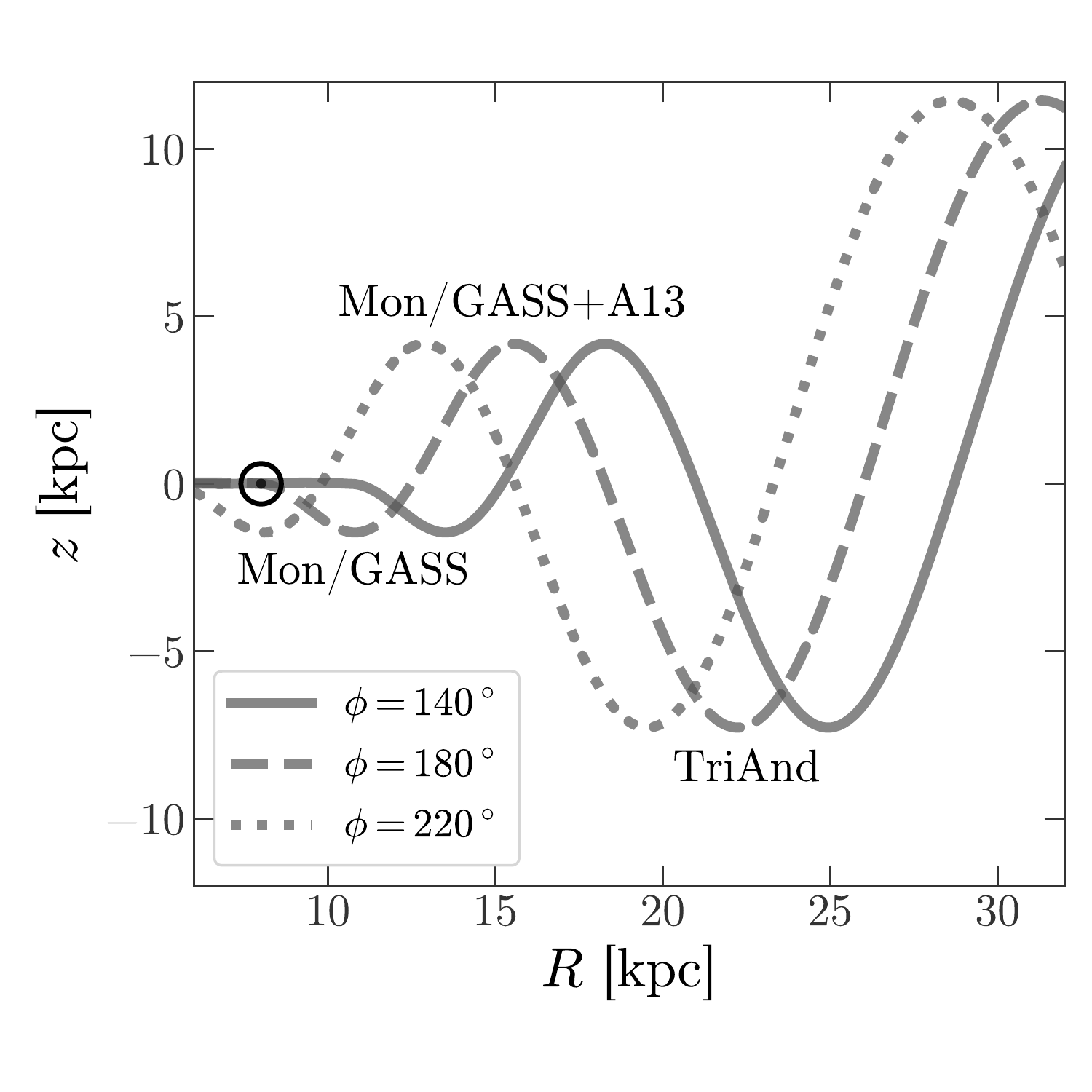}
\caption{Schematic of the distribution of Mon/GASS, Mon/GASS+A13, and the TriAnd cloud in the $R$-$z$ plane if the location of these stellar features is due to the vertical oscillation in the Galactic disk. The solid, dashed, and dotted curves show three different lines of sight; $\phi$ is a Galactocentric azimuth angle and increases in the opposite sense as the heliocentric Galactic longitude, and $R$ is the Galactocentric distance.}
\label{rzlos}
\end{centering}
\end{figure}

The Galactic disk exhibits a vertical asymmetry in the density of stars \citep{widrow12, slater14} and bulk motions of stars \citep{carlin13, williams13}, and this asymmetry is oscillating \citep{xu15}. 
These results support the conclusions of \citet{weinberg91} and \citet{widrow12}, suggesting that bending and breathing modes in the disk should be observed both locally and, at larger amplitudes, at higher $Z$.
The disk was likely set in motion by a fairly massive (e.g., Sgr, LMC) satellite galaxy plunging through the MW's disk. Concentric arc-like densities of stars that increase in scale height with distance are set in motion at the satellite's passage through the MW's disk, causing ripples in the disk. Simulations have shown that a satellite piercing the Galactic midplane can create the vertical asymmetries seen locally \citep{widrow12} and on the outer fringes of the disk \citep{donghia16}, at the disk-halo interface. Specifically, Sgr has been shown in N-body simulations to create such vertical oscillations (\citealt{purcell11}, \citealt{gomez13}, \citealt{laporte17}).

Although the distance estimates are rough and were derived using spectral indices \citep{sheffield14,li17}, the spatial distribution of the outer disk structures Mon/GASS, A13, and TriAnd also suggests that they could be part of a ripple in the disk in support of the \citep{xu15} finding. An illustration of this scenario is shown in Figure \ref{rzlos}, where the stellar overdensities will manifest along different azimuthal lines of sight, owing to their continuity in Galactic longitude. The Northern portion of Mon/GASS and A13 will appear at peak amplitude at $\phi$ = 180$^{\rm o}$, while the TriAnd cloud has a peak around $\phi$ = 140$^{\rm o}$, where the densest regions of the cloud are found \citep{ibata17}. In the schematic, the outwardly propagating material is likely more complex (e.g., spirals) than outwardly propagating rings at constant Galactocentric radii.  

\subsection{\label{sims}Consistently Simulating the Stellar Structures}
\begin{figure*}
\begin{centering}
\includegraphics[scale=0.4]{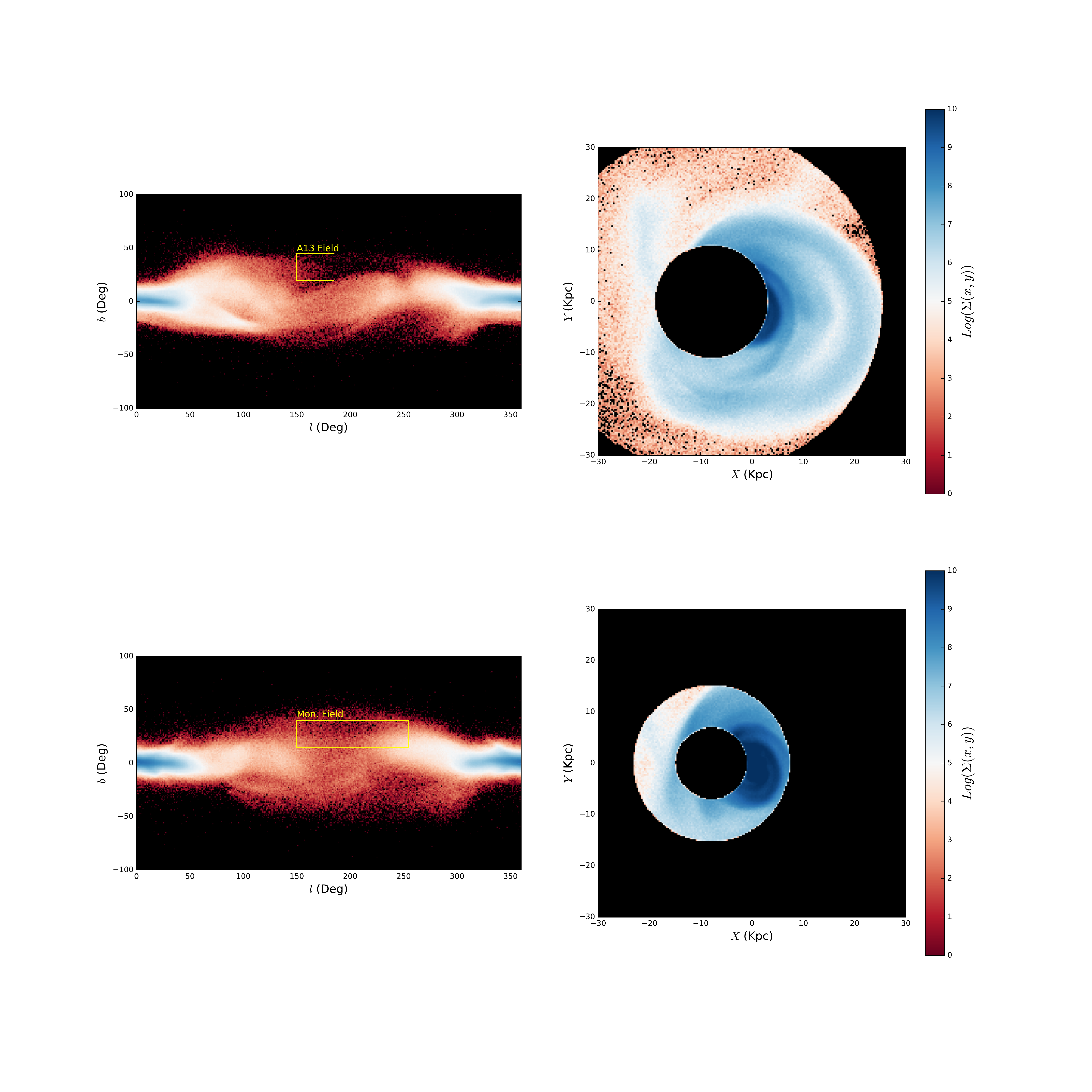}
\caption{Log number density plots at 5.6 Gyr (approximate present day position of the Sgr dSph galaxy). The top and bottom left panels are in observational space, restricted by annuli at 11-33 kpc for A13 and 7-15 kpc for Mon/GASS. On the right, the annuli are shown in $X$-$Y$ space.}
\label{dens}
\end{centering}
\end{figure*}

To gain further perspective on the connection between Mon/GASS and A13, N-body simulations were run to assess the impact on the MW's disk by a Sgr-type progenitor, with an initial mass of 10$^{11}$ M$_{\sun}$ \citep{laporte17}. 
The simulation takes into account the orbit of the satellite as it enters the virial radius of the MW and follows its evolution for a time of $\sim5.6\, \rm{Gyr}$ until the present-day. During its lifetime, Sgr makes 5 pericentric passages which excite vertical density waves, gradually kicking disk stars to higher latitudes. Because this model includes previous pericentric passages which were ignored in all previous studies \citep{purcell11, gomez13, laporte18}, stars far out in the disk are able to reach extreme heights about the midplane with $|Z|\sim10 \,\rm{kpc}$, well into the regions where A13 and TriAnd are observed. During its last pericentric passage, Sgr excites the Monoceros Ring which extends above and below $b\sim30^{\circ}$. In Figure \ref{dens} we show the distribution in $(l,b)$ of the present-day stars in the simulation within a heliocentric distance cut corresponding to the radii where the A13 and Monoceros regions are observed. We also mark the positions in the sky where our sample stars have been selected. We note that these fall within the expected heights for kicked-out disk stars in the MW following the interaction with Sgr.

The simulations of \citep{purcell11, gomez13, laporte18} have caveats, in that the vertical oscillations were not to scale. The simulations of \citet{laporte17} show that there is a model in which the locations of all three stellar features are explained by the interaction of Sgr with the Galactic disk. This complements the work of \citet{deason17}, which shows a relatively quiet accretion history of the MW's halo.

\section{\label{summary}Summary}
In this work, we present the results of a spectroscopic study designed to study two low-latitude stellar overdensities: A13 and Mon/GASS. Using stellar populations, we examined the origin of these overdensities and our results support a disk origin for both A13 and Mon/GASS. 

M giants in Mon/GASS and A13 have a low velocity dispersion and display a gradient as a function of Galactic longitude \citep{li17}, whereas we find the velocity dispersion of the RR Lyrae stars in both Mon/GASS and A13 is much higher and is consistent with halo membership. Using the same statistical approach as PW15 in their study of the TriAnd cloud, the fraction of RR Lyrae stars to M giants in Mon/GASS and A13 was found to be consistent with the Milky Way's disk, rather than an accreted satellite. An accreted population will have few (if any) M giant stars but many RR Lyrae stars, as the gas that forms stars in satellite galaxies it too metal poor to form M giants, the exception being luminous satellites such as Sagittarius.  

Prior connections between TriAnd and Mon/GASS were found by \citet{xu15}, who spatially analyzed stars in these two features. \citet{li17} analyzed the velocity distributions of M giants in Mon/GASS, A13, and TriAnd and found contiguous velocity sequences and low velocity dispersions. TriAnd also has a low fraction of RR Lyrae stars to M giants (PW15). Lastly, high-resolution chemical abundances lend direct support to a disk origin for TriAnd and A13 (Bergemann et al. 2017). Our work bolsters the associations between these 3 stellar features, whereby stars in Mon/GASS, A13, and TriAnd are all connected and formed from the same gaseous material in the Milky Way's disk.

\acknowledgements
We thank the anonymous referee for his/her helpful comments and suggestions. A.A.S. acknowledges support for this project provided by a Cycle 48 PSC-CUNY Research Award, jointly funded by The Professional Staff Congress and The City University of New York, and the CUNY Research Scholars Program. A.A.S. thanks John Thorstensen and Jeffrey Carlin for useful conversations. We thank Emily Sandford for assistance with observations. K.V.J. acknowledges NSF grants AST-1312196 and AST-1614743. C.L. is supported by a Junior Fellow of the Simons Society of Fellows award from the Simons Foundation. This work used the Extreme Science and Engineering Discovery Environment (XSEDE), which is supported by National Science Foundation grant number OCI-1053575. This work is based on observations obtained at the MDM Observatory, operated by Dartmouth College, Columbia University, Ohio State University, Ohio University, and the University of Michigan.

\end{document}